\documentclass[a4paper]{jpconf}
\usepackage{graphicx}
\usepackage{amssymb}
\usepackage{amsmath}
\usepackage{amsfonts}

\newcommand{\degr}{^\circ}
\newcommand{\rd}{\mathrm{d}}
\newcommand{\pc}{\mathrm{pc}}
\newcommand{\cnd}{\mathrm{CND}}
\newcommand{\bn}{\mathbf{n}}
\newcommand{\msun}{M_\odot}
\newcommand{\mbh}{M_\bullet}
\newcommand{\mcnd}{M_\mathrm{CND}}
\newcommand{\rcnd}{R_\mathrm{CND}}
\newcommand{\sgra}{SgrA$^\ast$}
\newcommand{\bsp}{\!\!\!}

\begin{document}
\title{Warping the young stellar disc in the Galactic Centre}

\author{Ladislav \v{S}ubr$^{1,2}$ and Jaroslava Schovancov\'a$^1$}

\address{$^1$Faculty of Mathematics and Physics, Charles University,
V Hole\v{s}ovi\v{c}k\'ach 2, CZ-18000 Praha, Czech Republic}
\address{$^2$Astronomical Institute, Academy of Sciences,
Bo\v{c}n\'{\i}~II, CZ-14131~Praha,
Czech Republic}

\ead{subr@sirrah.troja.mff.cuni.cz}

\begin{abstract}
We examine influence of the circum-nuclear disc (CND) upon the orbital
evolution of young stars in the Galactic Centre. We show that gravity of
the CND causes precession of the orbits which is highly sensitive upon
the semi-major axis and inclination. We consider such a differential
precession within the context of an ongoing discussion about the origin
of the young stars and suggest a possibility that all of them have originated
in a thin disc which was partially destroyed due to the influence of the CND
during the period of $\sim6\mathrm{Myr}$.
\end{abstract}

\section{Introduction}
Recent observations of the central stellar cusp of the Milky Way have revealed
a numerous population of massive young stars within the (projected) distance
of $\lesssim0.5\mathrm{pc}$ from the supermassive black hole (e.g. Genzel
et al.~1996, Paumard et al.~2006). The apparent youth of these
stars is considered to be in contradiction with their vicinity to the
black hole which is assumed to prevent stellar formation due to strong
tidal forces acting on the parent gaseous clouds. Further analyses have
shown that subset of stars above the projected radius $r\approx0.04\pc$
forms a coherently rotating `clockwise' disc (CWS). It has been proposed that
these stars may have been born in a massive gaseous disc fragmenting due
to it own gravity (Levin \& Beloborodov~2003, Nayakshin~2006). In spite of
its attractivity, this model, cannot explain origin of all young stars in
the central parsec as more than one half of them have large inclinations with
respect to the CWS. Even if we admit existence of another `counter-clockwise'
disc of $\lesssim 20$ young stars (Genzel et al.~2003, Paumard et al.~2006)
which could
also have been born via fragmentation of a gaseous disc, we will still be
facing a problem of the origin of several tens of massive young stars that
apparently do not belong to any disc-like structure in the Galactic Centre.

The aim of this contribution is to discuss dynamical processes that should
be taken into considerations about the relation of the initial and current
kinematical states of the young stars. In the following section, we present
notes on the dynamics of stars in the idealised model of the Galactic
Centre. In Section~\ref{sec:results} we discuss consequences of the orbital
evolution of the young stars with regard to the stability of a disc-like
stellar system. We summarise our results in Section~\ref{sec:conclusions}.

\section{Stellar dynamics in the central parsec}
\begin{figure}[t]
\begin{center}
\includegraphics[width=\textwidth]{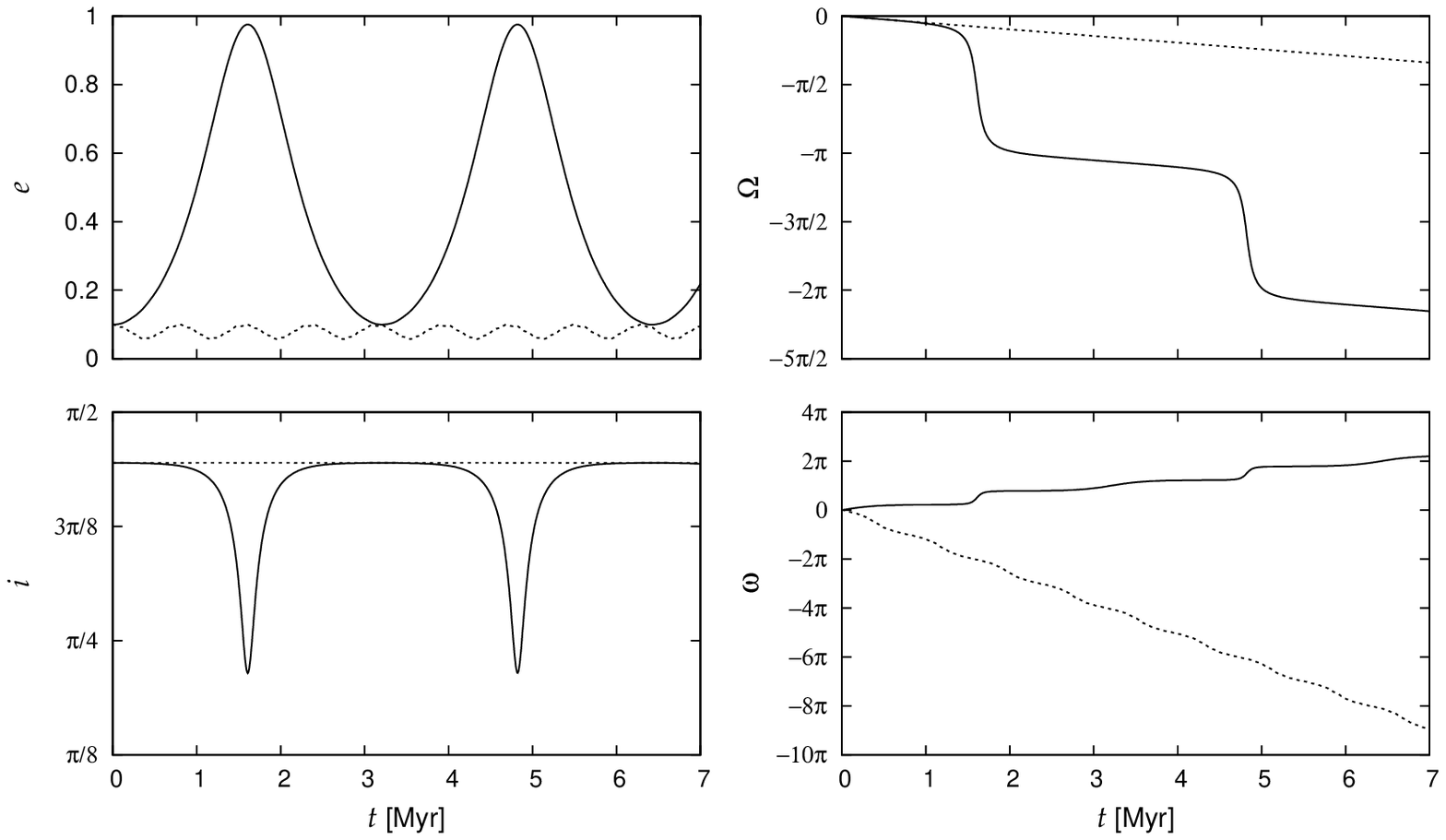}
\end{center}
\caption{Evolution of the orbital elements of two example orbits. The solid
line represents a trajectory in the gravitational field of the central mass
$\mbh=3.5\times10^6\msun$ and a ring of radius $\rcnd=1.5\mathrm{pc}$
and mass $\mcnd=\mbh$. The dotted line shows an orbit
integrated in the field including in addition a spherical cusp of mass
$M_\mathrm{c}=0.1\mbh$. In both cases the initial values of the orbital
elements are: $a=0.1\rcnd,\, e=0.1,\, i=80\degr,\, \omega=0$ and $\Omega=0$.}
\label{fig:kozai}
\end{figure}

Gravitational field within the central parsec is dominated by the supermassive
black hole. Relativistic effects can be safely ignored at the radii of
$r \gtrsim0.04\pc$ and, therefore, we will treat the black hole as a source
of purely Keplerian potential. In spite of the extreme stellar density in
this region,
characteristic time-scale of two-body relaxational processes are longer than
the lifetime of the young stars (Hopman \& Alexander~2006). Therefore, we will
model the gravitational
field of the stellar cusp with a smooth spherically symmetric potential
\begin{equation}
V_\mathrm{c}(r) = \frac{4GM_\mathrm{c}}{R_\mathrm{h}}
\left( \frac{r}{R_\mathrm{h}} \right)^{1/4}
\end{equation}
where $M_\mathrm{c}$ is the mass of the star cluster within the radius of the
influence of the black hole, $R_\mathrm{h}\approx1.5\pc$. The effect of the
stellar cusp upon the dynamics of the individual stars lies in a (negative)
advance of the pericentre of their orbits. Finally, we assume that a
non-spherical
component of the gravitational field is determined by the massive molecular
torus. For the sake of simplicity we will treat it as an infinitesimally
narrow ring of mass $M_\cnd$ and radius $R_\cnd$. Dynamics of the
stellar orbit
is then equivalent to the
dynamics in a reduced hierarchical triple system. Hence, we may describe
secular evolution of the orbital elements with equations
(Kozai~1962, Lidov~1962):
\begin{eqnarray}
T_\mathrm{K}\,\sqrt{1-e^2}\,\,\frac{\rd e}{\rd t} &\bsp=\bsp&
 {\frac{15}{8}}\,e\,(1-e^2)\,\sin2\omega\,\sin^{2}i\,,
 \label{eq:dedt} \\
T_\mathrm{K}\,\sqrt{1-e^2}\,\,\frac{\rd i}{\rd t} &\bsp=\bsp&
 -\frac{15}{8}\,e^2\,\sin2\omega\,\sin i\,\cos i\,,
 \label{eq:didt} \\
T_\mathrm{K}\,\sqrt{1-e^2}\,\,\frac{\rd\omega}{\rd t} &\bsp=\bsp&
 \frac{3}{4}\left\{ 2-2e^2+5\sin^{2}\omega\left[e^{2}-\sin^{2}i\right]\right\}\,,
\label{eq:dodt} \\
T_\mathrm{K}\,\sqrt{1-e^2}\,\,\frac{\rd\Omega}{\rd t} &\bsp=\bsp&
 -\frac{3}{4}\cos i \left[1+4e^2-5e^2\cos^{2}\omega\right]\,,
\label{eq:dOdt}
\end{eqnarray}
where
\begin{equation}
T_\mathrm{K} \equiv\frac{\mbh}{\mcnd} \, \frac{\rcnd^3}{a\sqrt{G\mbh a}}\;.
\label{eq:TK}
\end{equation}
is the characteristic time-scale of the orbital evolution, $a$ is the semi-major
axis of the orbit, $e$ is its eccentricity, $i$ is inclination with respect
to the CND, $\omega$ is the argument of the pericentre and $\Omega$ stands for
the longitude of the ascending node (all angles are measured in the reference
frame in which CND lies in the equatorial plane). Due to the
presence of two integrals of equations (\ref{eq:dedt}) -- (\ref{eq:dOdt})
and periodic nature of the angles $\omega$ and $\Omega$, inclination and
eccentricity of the orbit periodically oscillate. Example of these so called
`Kozai oscillations' is presented in Fig.~\ref{fig:kozai} (solid line).

Gravity of the stellar cusp can be included by means of an additional term
to equation~(\ref{eq:dodt}) which describes the additional pericentre
shift (Ivanov, Polnarev \& Saha~2005). In spite of that its explicit form
is not known for the arbitrary profile of $V_\mathrm{c}(r)$, we may safely
state that, in the absolute value, it overwhelms the pericentre advance due
to the CND already for $M_\mathrm{c} \gtrsim 0.1M_\cnd$. The periodic nature
of the changes of eccentricity and inclination is preserved. Nevertheless, its
period is shortened and the amplitude is substantially damped (Karas \&
\v{S}ubr~2007). As in the case of the Galactic Centre the mass of the star
cluster is well above the threshold value, we may omit the discussion
of the detailed profile of the stellar cusp. We may also consider secular
evolution of the eccentricity and inclination to be negligible.
Formula~(\ref{eq:dOdt}) for the precession rate can be then simplified by
averaging over one period of $\omega$ which yields:
\begin{eqnarray}
 \Delta \Omega &\bsp=\bsp& - \frac{3}{4}\, \cos i\; a^{3/2}\,
 \frac{\sqrt{G\mbh}}{\rcnd^3} \frac{\mcnd}{\mbh}\,
 \frac{1+\frac{3}{2}e^2}{\sqrt{1-e^2}}\, \Delta t \nonumber \\
 &\bsp=\bsp& -5417\degr\, \cos i\, \left( \frac{a}{\rcnd} \right)^{3/2}
 \left( \frac{\mbh}{3.5\!\times\! 10^6\msun} \right)^{1/2}
 \left( \frac{\rcnd}{1\mathrm{pc}} \right)^{-3/2}
 \frac{1+\frac{3}{2}e^2}{\sqrt{1-e^2}}\, \frac{\mcnd}{\mbh}\,
 \frac{\Delta t}{1\mathrm{Myr}}\;
 \label{eq:dO}
\end{eqnarray}

\begin{figure}[t]
\begin{center}
\includegraphics[width=0.75\textwidth]{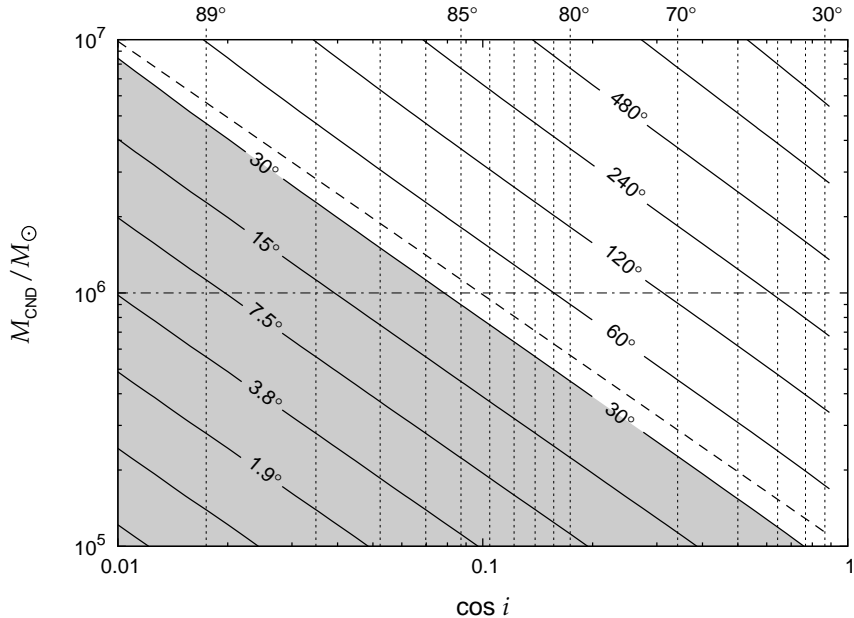}
\end{center}
\caption{$\Delta\Omega$ as a function of the mass of the CND and the mutual
inclination of the CND and a reference stellar orbit which evolves for a
period of $6\mathrm{Myr}$. Fixed is the radius of the CND, $R_\cnd=1.5\pc$,
semi-major axis of the orbit, $a=0.15\pc$, and its eccentricity, $e=0.5$.
Solid lines represent iso-contours of $\Delta\Omega$ as obtained by means of
the numerical integration of the orbit. Dashed line shows iso-contour of
$\Delta\Omega = 30\degr$ determined by formula~(\ref{eq:dO}).}
\label{fig:dOmega}
\end{figure}

We have performed several numerical tests of the validity of formula~(\ref{eq:dO}).
Figure~\ref{fig:dOmega} shows iso-contours of $\Delta\Omega$ as a function of
$M_\cnd$ and $\cos i$ obtained by means of numerical integration of the test
particle trajectory. The slope of the curves follows the theoretically predicted
one. The systematic shift of the iso-contours with respect to the values given
by eq.~(\ref{eq:dO}) as shown for $\Delta\Omega = 30\degr$ is mainly due to an
inconsistence between the mean and osculating values of the orbital elements.
Eqs.~(\ref{eq:dedt}) -- (\ref{eq:dO}) are valid for orbital elements averaged
over one revolution; on the other hand, $a=0.15\pc$ stands as an initial value
of the osculating semi-major axis in the numerical integration, while the mean
value of semi-major axis is slightly smaller due to the mass of the stellar
cusp. We haven't made correction to this discrepancy as it is not
straightforward and the introduced error of $\Delta\Omega$ is smaller than
$20\%$.

\section{Results}
\label{sec:results}
An important consequence of formula (\ref{eq:dO}) is a strong dependence of the
precession rate upon the semi-major axis and inclination. In the latter case,
the dependence is pronounced for $i\approx 90\degr$. For example,
the absolute value of the change of $\Omega$
for $a=0.4\mathrm{pc},\, e=0.5,\, i=85\degr,\, \rcnd=1.5\mathrm{pc},\
\mcnd=10^6\msun$ and $\Delta t=6\mathrm{Myr}$ is $|\Delta\Omega| \approx
100\degr$. On the other hand, for $a=0.04\mathrm{pc}$ and $i=89\degr$ and
keeping the other parameters unchanged, we obtain $|\Delta\Omega| \approx
1\degr$. Note that, if both orbits start with identical value of $\Omega$,
they would be {\em initially\/} corotating. These considerations
indicate that a popular model of the origin
of the young stars in the Galactic Centre which assumes their formation in a
fragmenting gaseous disc needs to include gravitational influence of the CND in
order to follow further evolution of the initially flat stellar disc.

\subsection{Orientation and mass of the CND}
Properties of the molecular torus are determined from radio observations of
emission of various molecules, both neutral and ionised. The maximum of
the observed emission is at a distance of $\sim1.5\mathrm{pc}$ from \sgra.
Its total mass is rather uncertain --- values ranging from $2\times10^5\msun$
to $2\times10^6\msun$ can be found in the literature. The most recent works
prefer higher masses (e.g. $M_\cnd \approx 10^6\msun$, Christopher
et al.~2005). Inclination with respect to the plane of the sky
is $i^\prime_\cnd \approx 70\degr$ and the position of the ascending (receding)
node measured from the North is $\Omega^\prime_\cnd \approx 25\degr$
(Jackson et al.~1993). (Note that angles $i$ and $\Omega$, i.e. without primes,
used in the previous section are measured in the coordinate system in which
the CND lies in the $z=0$ plane.) Uncertainty in the values of both angles
determining orientation of the CND is larger than $10\degr$.

Presence of the coherently rotating disc of young stars poses some constraints
upon the parameters of the CND. Assumption that the stellar disc was stable
for the period of $\sim6\mathrm{Myr}$ determines an upper limit of the rate of
the precession of its members --- the change of $\Omega$ must not exceed the
current value of the opening angle of the CWS which is $\approx 20\degr$
(Paumard et al.~2006). Figure~\ref{fig:dOmega} shows $\Delta\Omega$ as a
function of $M_\cnd$ and $i$ for an orbit at the outer edge of the
stellar disc. A particular diagonal line which represents the iso-contour of
$\Delta\Omega = 30\degr$ in $6\mathrm{Myr}$ can be treated as an boundary
of the (shaded) region of the values of $M_\cnd$ and $i$ that would not lead
to destruction of the CWS during its lifetime. Thin dash-dotted line which
indicates the estimated mass of the CND crosses the boundary at
$i\approx85\degr$. This limiting mutual inclination of the stellar disc and
the CND also determines an upper limit for the initial opening angle of the
stellar disc to $|90\degr - i_0| \approx 5\degr$.

\subsection{Single warped disc}
The values of parameters that lead a to non-destructive precession represent
rather small subset of their whole definition range. Hence, it is likely that
some stars were out of this region initially and, subsequently, they were
detached from the initial planar structure due to precession larger than
$20\degr$. Such a scenario could give an answer to the intriguing question
about the origin of those young stars in the Galactic Centre which apparently
do not belong to any of the recognised stellar discs.

\begin{figure}[t]
\begin{center}
\includegraphics[width=0.75\textwidth]{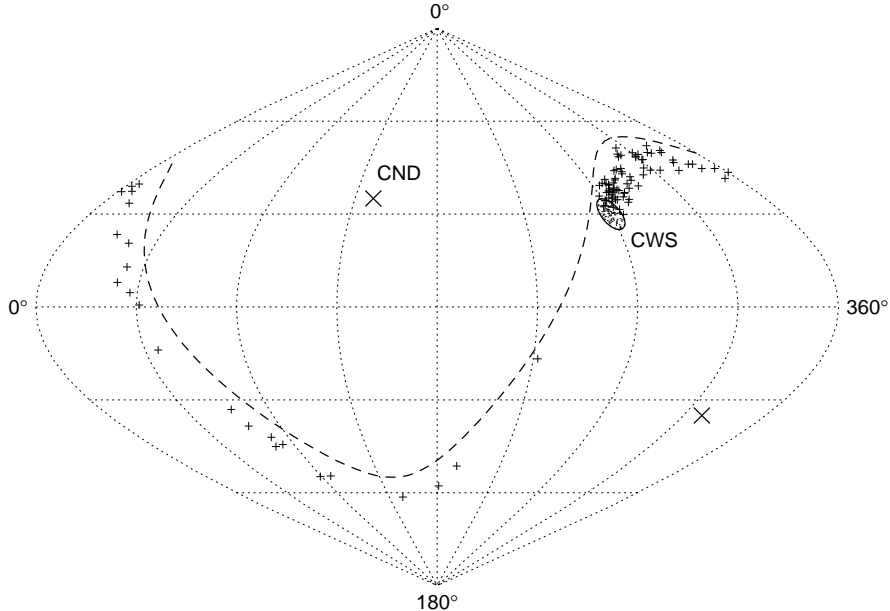}
\end{center}
\caption{Model of the warped disc in terms of sinusoidal projection of normal
vectors (small crosses) of individual orbits in the observer's reference frame.
Latitude represents angular
distance from the line of sight, longitude is measured counterclockwise from
the North. Large crosses denote position of the vectors normal to the CND;
dashed curve is a set of all vectors perpendicular to the axis of the CND.
The orbits have evolved under the influence of the CND for a period of
$6\mathrm{Myr}$, all of them originating from a small region around the
point $(\varphi^\prime_0,\,\theta^\prime_0) = (270\degr,\, 60\degr)$ the
boundary of which is represented by solid line.}
\label{fig:warp}
\end{figure}
The high sensitivity of the precession upon undetermined orbital parameters
of the observed stars as well as the lack of robust determination of the
mass and orientation of the CND prohibits us to track the stars back in time
and to seek for their parent disc(s).
We assume that forthcoming observations will enable us
to measure accelerations of the young stars in the Galactic Centre and,
consequently, to determine their orbital elements. Then, sets of stars that
were corotating in the past should be distinguished due to specific pattern
formed by their normal vectors. We show an example of the warped disc on
simulated data in Figure~\ref{fig:warp}. We plot
the configuration of normal vectors of 100 orbits that were detached from
a $5\degr$ neighbourhood of the vector $\bn_0 = (\varphi^\prime_0,\,
\theta^\prime_0) = (270\degr,\, 60\degr)$ which corresponds to
$(i^\prime_0,\,\Omega^\prime_0) =
(120\degr,\, 90\degr)$. The space occupied initially by the normal vectors
is denoted with a small circle (deformed due to the sinusoidal
projection), while their positions after $6\mathrm{Myr}$ of the orbital
evolution are denoted with crosses. Distribution of the semi-major axes is
$n(a)\propto a^{-1}$ which transforms to the surface density $\propto r^{-2}$;
eccentricities are drawn randomly with equal probability from an interval
$(0,1)$. Parameters of the CND were set to $M_\cnd = \mbh,\, R_\cnd = 1.5\pc$
and $(\varphi^\prime_\cnd,\,\theta^\prime_\cnd) =
(325\degr,\, 125\degr)$. Most of the orbits' normal vectors remained
relatively close to their original positions, forming a structure that can be
still recognised as a stellar disc. Those stars that have undergone more
rapid precession form a specific pattern in the space $(\varphi^\prime,
\theta^\prime)$ --- a tail that is parallel with the circumference perpendicular
to the CND.

In \v{S}ubr, Schovancov\'a \& Kroupa~(2009) we have performed a test of
compatibility of the hypothesis of a common origin of the young stars in the
GC in a single thin disc with the observational data published by Paumard et
al.~(2006, Table 2). We have examined a $1\sigma$ neighbourhood of the
velocities of 72 stars (excluding S-stars and stars with undetermined velocity
component along the line of sight) searching for such a state, initial
orientation of which is close to
a given normal vector $\bn_0$. We assumed the orbital evolution to be solely
due to the precession around the axis of the CND, orientation and mass of which
stand as free parameters. We have shown that there exists a class of the
parameters of the CND for which all stars might have a common origin in a thin
disc. Not surprisingly, the best fits gave initial orientation of the disc
around $(i^\prime_0,\,\Omega^\prime_0) = (120\degr,\, 90\degr)$, which is close
to the orientation of the clockwise stellar disc (Levin \& Beloborodov~2003,
Paumard et al.~2006). Hence, the hypothesis of the common origin of all young
stars in the Galactic Centre (except for S-stars) can be considered viable.

\section{Conclusions}
\label{sec:conclusions}
Within our idealised model, the orbits of stars in the central parsec of the
Galaxy precess around the symmetry axis of the massive molecular torus. We
have verified by means of numerical integration of the equations of motion
that this process is robust and it is not substantially modified when the
axial symmetry is broken and CND is represented by a finite number of discrete
particles. We estimate that relaxation processes act on the young stars on
time-scales
considerably longer than their presumed lifetime. Therefore, we suggest that
the precession due to the gravity of the CND played a dominant role in the
orbital evolution of these stars during the past $\sim6\mathrm{Myr}$.

The rate of precession is sensitive upon the semi-major axis of the orbit
and its inclination with respect to the CND. This process is of particular
importance for the dynamics of a stellar disc -- the differential precession
tends to destroy a disc-like system which consists of orbits within some range
of semi-major axis an inclination. As an example, we have shown that for a
system with $a \in (0.04\mathrm{pc},\, 0.4\mathrm{pc})$ and an opening angle
of $\approx 5\degr$ which is nearly perpendicular to the CND, the precession
will drag normal vectors of a considerable fraction of stars by more than
$20\degr$ away from their original positions.

We point out that, due to the differential precession, there is a nontrivial
mapping between initial and current orientations of orbits of the young
stars in the Galactic Centre. We suggest a partial destruction (warping)
due to the gravity of the massive gaseous torus of
an initially coherently rotating disc as
an explanation of the origin of those young stars in the Galactic Centre that
do not belong to the recently identified clockwise stellar disc. Our hypothesis
of a warped disc predicts that normal vectors of all of its member stars should
lie close to the circumference perpendicular to the axis of the CND.
We have verified that this hypothesis is compatible with the observational data
published by Paumard et al.~2006. Nevertheless, the lack of robust
determination of the current values of the orbital elements form the
observational data prevents a convincing identification of the warped stellar
disc. Finally, let us remark that the hypothesis of the single warped stellar
disc also cannot explain apparently random orientations of the orbits of the
S-stars, as these are the least affected by the gravity of the CND.

\ack
This work was supported by the Research Program MSM0021620860 of the Czech
Ministry of Education, the Centre for Theoretical Astrophysics in Prague
and the Czech Science Foundation (ref.\ 205/07/0052).


\section*{References}
\begin{thebibliography}{9}
\bibitem{christopher05}
 Christopher M~H, Scoville N~Z, Stolovy S~R, Yun M~S 2005 {\it ApJ} {\bf 622} 346
\bibitem{genzel96}
 Genzel R, Thatte N, Krabbe A, Kroker H, Tacconi-Garman L~E 1996 {\it ApJ}
 {\bf 472} 153
\bibitem{hopman06}
 Hopman C, Alexander T 2006 {\it ApJ} {\bf 645} 1152
\bibitem{ivanov05}
 Ivanov P~B, Polnarev A~G, Saha~P 2005 {\it MNRAS} {\bf 358} 1361
\bibitem{jackson93}
 Jackson J~M, Geis N, Genzel R, Harris A~I, Madden S, Poglitsch A,
 Stacey G~J, Townes C~H 1993 {\it ApJ} {\bf 402} 173
\bibitem{karas07}
 Karas~V, \v{S}ubr~L 2007 {\it A\&A} {\bf 470} 11
\bibitem{kozai62}
 Kozai~Y 1962 {\it AJ} {\bf 67} 591
\bibitem{levin03}
 Levin~Y, Beloborodov A~M 2003 {\it ApJ} {\bf 590} L33
\bibitem{lidov62}
 Lidov M~L 1962 {\it Planetary and Space Sci.} {\bf 9} 719
\bibitem{nayakshin06}
 Nayakshin S 2006 {\it MNRAS} {\bf 372} 143
\bibitem{paumard06}
 Paumard~T, Genzel~R, Martins~F et~al.  2006 {\it ApJ} {\bf 643} 1011
\bibitem{subr08}
 \v{S}ubr L, Schovancov\'a J, Kroupa P 2009 {\it A\&A} accepted
\end{thebibliography}
\end{document}